\documentclass[12pt]{iopart}
\usepackage{iopams}
\usepackage{amsfonts}
\usepackage{amssymb}
\usepackage{graphicx}
\usepackage{color}
\usepackage{bm}

\def\ketm#1{  \left\vert  #1   \right\rangle   }

\begin{document}

\title[Calculation of the two--photon decay rates of hydrogen--like ions]{Calculation of the two--photon decay rates of 
hydrogen--like ions by using B--polynomials}

\author{P. Amaro\dag $\|$,
A. Surzhykov\ddag \S,
F. Parente\dag,
P. Indelicato$\|$ and
J. P. Santos\dag}

\address{\dag\ CFA, Departamento de F{\'\i}sica, Faculdade de Ci{\^e}ncias e
  Tecnologia, \\
  FCT, Universidade Nova de Lisboa, 2829-516 Caparica, Portugal}

\address{\ddag\ Physikalisches Institut,
Universit\"at Heidelberg, Philosophenweg 12, D--69120 Heidelberg,
Germany}

\address{\S\ GSI Helmholtzzentrum f\"ur Schwerionenforschung,
  Planckstr. 1, D--64291 Darmstadt, Germany }

\address{$\|$\ Laboratoire Kastler Brossel, \'Ecole Normale
  Sup\'erieure, CNRS, Universit\'e P. et M. Curie -- Paris 6, Case 74;
  4, place Jussieu, 75252 Paris CEDEX 05, France}


\begin{abstract}

A new approach is laid out to investigate the two--photon atomic transitions. It is based on application of the finite basis solutions constructed from the Bernstein Polynomial (B--Polynomial) sets. We show that such an approach  provides a very promising route for the relativistic second- (and even higher-order) calculations since it allows for analytical evaluation of the involved matrices elements. In order to illustrate possible applications of the method and to verify its accuracy, detailed calculations are performed for the $2s_{1/2 }\rightarrow 1s_{1/2}$ transition in neutral hydrogen and hydrogen--like ions, and are compared with the theoretical predictions based on the well--established B--spline--basis--set approach. 

\end{abstract}


\pacs{31.15.-p, 31.30.Jv, 32.70.Fw, 32.80.Wr}

\submitto{\JPA}

\vspace{1cm}
\noindent {\bf \today }

\maketitle

%
\section{\label{intro}Introduction}

High--order perturbation calculations in atomic physics generally require summations over the \textit{complete} spectrum of 
the system under consideration. Within the relativistic framework, such a ``summation'' is not a simple task since it includes a 
summation over discrete part of
the Dirac spectrum as well as the integration over the positive-- as well as negative--energy continua. A number of methods 
have been developed over the past decades to perform this summation consistently. Apart from the various Green's function 
approaches ~\cite{3499,1353}, the \textit{discrete--basis--set} 
method is widely employed nowadays in (relativistic as well as non--relativistic) high--order calculations \cite
{3372,388,1077,SO_1989_5548, 
SO_1989_5559}. In this method, a \textit{finite} set of discrete pseudostates is constructed from some basis functions and 
utilized for carrying out the summation. The particular choice of a suitable set of basis functions is crucial for the practical implication 
of the method. Usually, the discrete (pseudo--) solutions are built up from \textit{piecewise polynomial} sets. 
The piecewise polynomials are precisely defined, can be calculated rapidly on modern computer systems, and can represent a great 
variety of functions. They can be differentiated and integrated easily~\cite{3379}.
%
%

The basis splines~\cite{11}, also called B--splines, are one of the most commonly used family of piecewise polynomials. 
These polynomials, which are well adapted to numerical tasks, have been successfully used in many atomic--physics studies. 
For example, Johnson and co--workers have applied the B--splines to the many--body perturbation theory~\cite{8,3368}. 
Froese Fischer and co-workers used it in (variational) Hartree--Fock calculations and continuum problems ~\cite{3369,3370}. Qiu and 
Froese Fischer introduced the integration by cell algorithm for Slater integrals in a B--spline basis obtaining an improved efficiency and 
accuracy over traditional methods ~\cite{3371, FFZ_2009}. Bhatti and co--workers ~\cite{3410} used similar techniques to find an 
approximate solution of a set of the non--homogeneous second--order differential equations and to obtain static polarizabilities of  
hydrogenic states. 
Indelicato and co--workers employed B--splines in the multi--configuration Dirac--Fock (MCDF) 
relativistic atomic structure calculations ~\cite{47,62} and in relativistic two--photon decay calculations ~\cite
{388,3007,3140}.

While B--spline basis sets were proven to be an important tool for studying the variety of atomic structure and dynamics 
problems (see Ref.~\cite{3372, FF_2008} for further details and examples),  one might adopt other piecewise polynomial sets to speed up relativistic high-order calculations as might be highly required, for example, for studying the two-photon transitions in many-electron systems.
%
%
In this work we argue 
that the Bernstein, or B--polynomials ~\cite{3373} may serve as a good alternative to the B--splines since they allow for \textit
{analytical} finite--basis--set calculations. These are polynomial functions of $n^{\mathrm{th}}$ degree that have been 
recently used to obtain the solution of some 
linear and non-linear differential 
equations~\cite{1452, 3363, 3364}. 
Bhatti and Perger~\cite{1449} developed an algorithm for constructing accurate solutions to the radial Dirac
equation in a B-polynomial basis set. 
By using this algorithm and the Galerkin method, accurate calculations have been performed for the bound-state energies of hydrogenic systems.



In this work, we employ the finite (discrete) solutions constructed from the B--polynomial sets in order to explore two--photon 
decay of hydrogen--like ions. Theoretical analysis of this process requires evaluation of the second--order transition 
amplitudes and, hence, can be used as a ``testing ground'' for the high--order B--polynomial calculations. To explain the 
background of these calculations we will recall in Sections \ref{Dirac_equation} and \ref{Boundary_conditions} the 
application of finite--basis--set methods for dealing with the Dirac problem. In particular, we will derive the generalized 
eigenvalue problem whose solutions form a complete set of atomic pseudostates. In order to find these solutions, however, 
one has to agree first about the explicit form of the basis functions. In Section~\ref{Finite_basis_expansion}, we 
will introduce  B--polynomial basis sets, and discuss their properties.  
We will use these sets to construct the (pseudo) spectrum of the ion (or atom). Summation over such a spectrum, which appears within the second--order perturbation theory, will be performed later in Section \ref{Dif_total_decay_rates} and will allow us to derive the two--photon decay rates. Results of the 
relativistic calculations for these rates obtained for the $2s_{1/2 }\rightarrow 1s_{1/2}$ transition in neutral hydrogen and 
hydrogen--like ions will be presented in Section \ref{Results_discussion}. Apart from the results obtained by making use of 
the B--polynomial sets, we present here also the ``standard'' B--spline calculations as well as the predictions by Labzowsky 
\etal ~\cite{1418}. Detailed comparison with these predictions will allow us to justify the application of the B--polynomial 
aproach in second--order calculations and to underline its advantages. The summary of our work will be given finally in 
Section~\ref{Conclusions}. 

The atomic system of units ($e$ = $m_e$ = $\hbar$ = 1) is used throughout the paper unless otherwise stated.
%
%
%
\section{Theory}
\label{theory}
%
%
\subsection{General approach}
{\label{General_approach}

The second--order relativistic calculations in atomic and molecular physics 
often require summation over the \textit{complete} spectrum of the system under consideration. Such a summation, that includes integration over the positive-- as well as negative--energy continua, can be performed very efficiently if one considers the (atomic or molecular) system to be enclosed in a finite cavity with a radius R. This allows for a discretization of the continua and, hence, for a representation of the entire Dirac spectrum in terms of the pseudostate basis functions. A (quasi--complete) finite set of these functions are determined subsequently by making use of the variational Galerkin method~\cite{179}.


%
%
\subsection{Finite basis set approach to the Dirac equation}
\label{Dirac_equation}

Having briefly discussed the general context, we are ready now to apply the finite basis set method for solving the 
eigenvalue Dirac's problem
\begin{equation}
   \label{eigenproblem}
   \left( c \boldsymbol{\alpha} \cdot \mathbf{p} + \beta  c^{2} + V(r) \right) u( \mathbf{r})= \varepsilon u( \mathbf{r}) \, ,
\end{equation}
where $\boldsymbol{\alpha}$ and $\beta$ are the usual $4 \times 4$ Dirac matrices, and $V(r)$ describes Coulomb 
interaction between an electron and nucleus. Moreover, in Eq.~(\ref{eigenproblem}) we have replaced the total electron 
energy $E$ by $\varepsilon = E -  c^{2}$ to render easy the comparison with nonrelativistic calculations. 

Since the potential $V(r)$ is central, the eigenfunctions of the Dirac Hamiltonian can be written in the standard form as  
%
%
\begin{equation}
  \label{dirac_wavefunction}
  u_{n \kappa}( \mathbf{r}) = \frac{1}{r}
  \left[
    \begin{array}{cc}
        P_{n \kappa}(r) & \Omega_{ \kappa m}( \hat{\mathbf{r}}) \\
      i Q_{n \kappa}(r) & \Omega_{-\kappa m}( \hat{\mathbf{r}})
    \end{array}
  \right] ,
\end{equation}
with $\Omega_{ \kappa m}( \hat{\mathbf{r}})$ being the Dirac spin--angular function.
The angular quantum number $\kappa$ is defined by
\begin{equation}
  \kappa= \left\{\begin{array}{lcl} \ell \quad&\mbox{if }&j=\ell-1/2\\
      -(\ell +1)\quad&\mbox{if
      }&j=\ell +1/2\end{array}\right. \ ,
\label{pkappa}
\end{equation}
where $\ell$ and $j$ are the electron orbital and total angular momenta, respectively. By substituting the wavefunction (\ref
{dirac_wavefunction}) into Eq.~(\ref{eigenproblem}) and by performing some simple angular momentum algebra one can 
obtain the conventional set of radial Dirac equations
\begin{eqnarray}
  \left[
    \begin{array}{cc}
      V(r) & c\  O_{-}^{\kappa} \\
      -c\ O_{+}^{\kappa}   & -2m_{e}c^{2}+V(r)
    \end{array}
  \right] & &
  \left[
    \begin{array}{c}
      P_{n \kappa}(r) \\
      Q_{n \kappa}(r)
    \end{array}
  \right]  
  = \varepsilon \left[
    \begin{array}{c}
      P_{n \kappa}(r) \\
      Q_{n \kappa}(r)
    \end{array}
  \right] .
\label{matrix_equation}
\end{eqnarray}
to determine the large, $P_{n \kappa}(r)$, and the small, $Q_{n \kappa}(r)$, radial components. For the sake of shortness, 
we introduced here the operator
\begin{equation}
  O_{\pm}^{\kappa} = \frac{d}{dr} \pm \frac{\kappa }{r}.
  \label{operator_o}
\end{equation}

For the further evaluation of Eq.~(\ref{matrix_equation}) we assume the ion (or atom) is enclosed in a finite cavity with a 
radius $R$ large enough to get a good approximation of the wavefunctions, with some suitable set of boundary conditions. 
In order to construct these functions, we shall turn to the principle of least action \cite{8}
\begin{equation}
   \label{least_action_principle} 
   \delta S_{\kappa}=0 \, ,
\end{equation}
from which the Dirac equation can be derived. In this expression, the action $S_{\kappa}$ is defined as 
%
%
\begin{eqnarray}
  S_{\kappa} &=&  \frac{1}{2} \int_{0}^{R}\left\{ c P_{n\kappa }(r)
O_{-}^{\kappa}
Q_{n\kappa }(r)\right.
  -c Q_{n\kappa }(r)
O_{+}^{\kappa}
P_{n\kappa }(r)  \nonumber \\
 & & + V(r)\left[ P_{n\kappa }(r)^{2}+Q_{n\kappa
    }(r)^{2}\right]
   \left. -2m_{e}c^{2}Q_{n\kappa }(r)^{2}\right\} dr  \nonumber \\
& & -\frac{1}{2} \, \epsilon \,
  \int_{0}^{R}\left[ P_{n\kappa }(r)^{2}+Q_{n\kappa }(r)^{2}\right] dr
 +  S_{\kappa}^{\mathrm{bond}}
  ,
  \label{rde1}
\end{eqnarray}
%
%
where the upper integration limit $R$ is the radius of the confining cavity. The term $ S_{\kappa}^{\mathrm{bond}}$, that will 
be specified below, stands for the boundary conditions and the parameter $\epsilon $ is a Lagrange multiplier introduced to 
ensure the normalization constraint,
\begin{equation}
  \int_{0}^{R}\left[ P_{n\kappa }(r)^{2}+
    Q_{n\kappa }(r)^{2}\right] dr=1 .
  \label{rde2}
\end{equation}
Here, the large, $P_{n\kappa }(r)$, and small, $Q_{n\kappa }(r)$, radial components of the electron wavefunctions can be 
written as a finite expansion 
\begin{eqnarray}
  P(r) &=&\sum_{i=1}^{N}p_{i}B_{i}(r),  \nonumber \\
  Q(r) &=&\sum_{i=1}^{N}q_{i}B_{i}(r),
  \label{rde4}
\end{eqnarray}
over some basis functions $B_{i}(r)$. The explicit form of these functions is not crucial for the following discussion and will 
be specified later in Section \ref{B_spline expansion}. In Eq.~(\ref{rde4}), moreover, the subscripts $n$ and $\kappa $ have 
been omitted from the functions $P_{n\kappa}(r)$ and $Q_{n\kappa}(r)$ for the sake of notation simplicity.

By inserting now the radial components (\ref{rde4}) into the least action principle (\ref{least_action_principle}) and by 
evaluating the variation $S_{\kappa}$ with respect to change of expansion coefficients $p_{i}$ and $q_{i}$, we obtain the 
matrix equation 
\begin{equation}
  A v = \epsilon B v  ,
  \label{rde5}
\end{equation}
to determine the vector $ v= \left(p_{1},p_{2},\ldots ,p_{N}, q_{1},q_{2},\ldots ,q_{N} \right)$ and where $A$ and $B$ are 
symmetric $2N \times 2N$ 
matrices given, respectively, by
%
\begin{equation}
  A=
  \left[
    \begin{array}{cc}
      (V) & c\left[ (D)- \left( \displaystyle\frac{\kappa }{r} \right)
      \right]  \\
      -c\left[ (D)+ \left( \displaystyle\frac{\kappa }{r} \right)
      \right]  & -2m_{e}c^{2}(C)+(V)
    \end{array}
  \right]
  +  A^{\mathrm{bond}}
  \label{rde7}
\end{equation}
%
and
\begin{equation}
  B=
  \left[
    \begin{array}{cc}
      (C) & 0 \\
      0 & (C)
    \end{array}
  \right] .
  \label{rde8}
\end{equation}
The matrix $A^{\mathrm{bond}}$ reflects the boundary conditions, and the $N\times N$ matrices $(C)$, $(D)$, $(V)$ and $
(\kappa /r)$ are given by
\begin{equation}
  (C)_{ij}=\int B_{i}(r)B_{j}(r)dr,
  \label{rde10a}
\end{equation}
\begin{equation}
  (D)_{ij}=\int B_{i}(r)\frac{d}{dr} B_{j}(r) dr,
  \label{rde10b}
\end{equation}
\begin{equation}
  \left( \frac{\kappa}{r} \right)_{ij}=\int B_{i}(r) \frac{\kappa}{r} B_{j}(r)dr.
  \label{rde10c}
\end{equation}
\begin{equation}
  (V)_{ij}=\int B_{i}(r) V(r) B_{j}(r) dr \, .
  \label{rde10d}
\end{equation}
%

Equation (\ref {rde5}) is known as a generalized eigenvalue problem that can be solved by employing linear algebra standard techniques. In the present work, for example, we have used the well-established LAPACK 3.3.0 package ~\cite{lapack}.
%
%
By using this package we obtain $2N$ real eigenvalues $\epsilon_{\lambda }$ and $2N$ orthogonal eigenvectors $v^{\lambda }$ that span both positive and negative energy solutions. Solutions labeled by $i=1,...,N$ describe the continuum $\varepsilon _{n} ^{i}<-2mc^{2}$ and solutions labeled by $i=N+1$,$..., 2N$ describe bound states and the positive continuum $\varepsilon _{n}^{i}>0$.

%
%
\subsection{Spurious states and boundary conditions}
\label{Boundary_conditions}

The practical implementation of the finite basis set approaches is usually complicated by the well--known problem of 
spurious states. These non--physical states appear as solutions of the single--particle radial Dirac equation for $\kappa >$ 0 
(p$_{1/2}$, d$_{3/2}$, $\ldots$ orbitals) ~\cite{3382}. Although the spurious solutions ``spoil'' the spectrum of the ion (or 
atom) under consideration, they are required for providing completeness of the basis set. The problem of spurious states 
has been discussed in detail in the literature, and several solutions were proposed ~\cite{3382, 3365,3381,3362,2987}. 
Johnson an co--workers ~\cite{8}, in their pioneering 
applications of the B--splines to the relativistic many--body problem, have suggested to adopt the function $S_{\kappa}^
{\mathrm{Bond}}$ in Eq. (\ref{rde1}) as
\begin{equation}
  S_{\kappa}^{\mathrm{Bond}} =
  \left\{
    \begin{array}{l}
      \frac{c}{4} \left[ P^{2}(R) -Q^{2} (R) \right] -\frac{c}{2} P(0)
      \left[ P(0) - Q(0) \right]  \\
      \hspace{4.5cm}\mathrm{for \ }  \kappa<0 \\
      \frac{c}{4} \left[ P^{2}(R) -Q^{2} (R) \right] -\frac{c}{2} P(0)
      \left[c P(0) - Q(0) \right]  \\
      \hspace{4.5cm}\mathrm{for \ }  \kappa>0 \\
    \end{array}
  \right. ,
\label{SBound_01}
\end{equation}
in order to \textit{lift} the spurious states to lower energies (to the negative continumm), thus restoring the low--energy mapping to the physical solutions. 
%
For variations of $P(r)$ and $Q(r)$ the boundary terms vanish if 
\begin{equation}
  P(0)=0 \hspace{0.5cm} \mathrm{and} \hspace{0.5cm} P(R)=Q(R).
\label{Bound_01}
\end{equation}
The latter boundary condition at the outer boundary $r=R$ is the MIT--bag-model condition~\cite{176}, and was included to avoid problems with Klein's paradox that arises when one attempts to confine a particle into a cavity, essentially by forcing the radial current crossing the boundary to vanish~\cite{132}.

%
%
%
\subsection{Finite-basis expansion}
\label{Finite_basis_expansion}

%
%
\subsubsection{B-polynomials expansion}
\label{B_poly expansion}

As seen from Eqs.~(\ref{rde5})--(\ref{rde10d}), any further analysis of the generalized eigenvalue problem requires the 
knowledge on the the explicit form of the basis functions $B_{i}(r)$. In the present work we will construct these functions from 
the B–-polynomial as well as B--spline spline sets. While the latter case will be discussed in Subsection \ref{B_spline 
expansion}, here we shall recall the main features of B--polynomials, also known as Bernstein functions.

The B--polynomials of $k$th--order are defined by~\cite{3373, 1452, 3363}
\begin{equation}
  B_{i,k}(r) = \left(\begin{array}{c} k \\ i \end{array} \right)
  \frac{\left( r-a\right) ^{i}\left( b-r\right)
    ^{k-i}}{\left( b-a\right) ^{k}} \, , \, \, \, \, i=0, 1, \ldots, k \, ,
\label{B_poly}
\end{equation}
where the standard form of the binomial coefficients are utilized, 
\begin{equation}
  \left(\begin{array}{c} k \\ i \end{array} \right) = \frac{k!}{i!(n-i)!},
\end{equation}
and where $a$ and $b$ denote the limits of the interval $\left[ a, b \right]$ over which the polynomials are defined to form a 
complete basis. Since the atomic system is defined in a finite cavity of radius $R$, we take $a=0$ and $b=R$.

As seen from definition (\ref{B_poly}), there are are $(k+1)$ polynomials of degree $k$. By definition, we set $B_{i,k}(r)=0$ if 
$i<0$ and $i>k$. 
%
As an example, a set of 11 B--polynomials of degree 10 is plotted in Fig.~\ref{fig_bpolys} where where it is shown that each B-polynomial is positive and the sum of all B-polynomials is unity.

%

The great advantage of B--polynomials is that they allow for an analytical evaluation of the matrices $(C)$, $(D)$ and $
(\kappa /r)$ in the generalized eigenvalue problem (\ref{rde5}). That is, by inserting $B_{i}=B_{i,k}(r)$ into Eqs.~(\ref
{rde10a})-(\ref{rde10c}), we obtain
\begin{eqnarray}
  (C)_{ij} &=&R
  \left(\begin{array}{c} k \\ i \end{array} \right)
  \left(\begin{array}{c} k \\ j \end{array} \right)
  \frac{1}{(2k+1)
    \left(\begin{array}{c} 2k \\ i+j \end{array} \right)
  } \, ,
  \label{C_pol} \\
  (D)_{ij} &=&
  \left(\begin{array}{c} k \\ i \end{array} \right)
  \left(\begin{array}{c} k \\ j \end{array} \right)
  \frac{(j-i)}{2(i+j)
    \left(\begin{array}{c} 2k-1 \\ i+j \end{array} \right) 
} \, ,
  \label{D_pol} \\
  \left( \frac{\kappa}{r}\right) _{ij} &=& \kappa
  \left(\begin{array}{c} k \\ i \end{array} \right)
  \left(\begin{array}{c} k \\ j \end{array} \right)
  \frac{1}{(i+j)
    \left(\begin{array}{c} 2k \\ i+j \end{array} \right) \, .
}
  \label{k/r_pol}
\end{eqnarray}
Apart from the basis functions $B_{i,k}(r)$, the knowledge on the nuclear charge distribution is also required for an 
evaluation of the matrix $(V)$ whose elements are defined by Eq.~(\ref{rde10d}). For the point--like nucleus potential, $ V^
{\mathrm{p}} (r)= -Z / r$, for example, these matrix elements read as
\begin{equation}
  (V^{\mathrm{p}})_{ij}  = -Z
  \left(\begin{array}{c} k \\ i \end{array} \right)
  \left(\begin{array}{c} k \\ j \end{array} \right)
  \frac{1}{(i+j)
    \left(\begin{array}{c} 2k \\ i+j \end{array} \right)
  } \, .
  \label{V_p_pol}
\end{equation}
A more complicated expression for the matrix $(V)$ is obtained to account for the finite nuclear size effects. To address these 
effects, we employ here the potential
\begin{equation}
V^{\mathrm{U}}(r)=\left\{
  \begin{array}{lll}
    \frac{Z}{2r_{\mathrm{N}}}\left[ \left( \frac{r}{r_{\mathrm{N}}}\right)
      ^{2}-3\right],& &    r\leq r_{\mathrm{N}} \\
    -\frac{Z}{r}, & & r>r_{\mathrm{N}}
  \end{array}
\right. ,
\end{equation}
due to a uniform spherical nuclear charge distribution with radius $r_{\mathrm{N}}$. By employing this potential in Eq.~(\ref
{rde10d}), we finally obtain
%
\begin{eqnarray}
  (V^{\mathrm{U}})_{ij}
  &=&(V^{\mathrm{p}})_{ij} + Z
  \left(\begin{array}{c} k \\ i \end{array} \right)
  \left(\begin{array}{c} k \\ j \end{array} \right)
  \left\{ B \left(\frac{r_{\mathrm{N}}}{R}; i+j,1-i-j+2k\right)
  \right.     \nonumber \\
  && \hspace*{-1cm} -\frac{1}{2}\left( \frac{r_{\mathrm{N}}}{R}\right) ^{i+j}\,\frac{3_{2}F_{1}\left[
      \left\{ 1+i+j,i+j-2k\right\} ,\left\{ 2+i+j\right\} ,\frac{r_{\mathrm{N}}}{R}\right]
  }{1+i+j}  \nonumber \\
  && \hspace*{-1cm} \left. +\frac{1}{2}\left( \frac{r_{\mathrm{N}}}{R}\right) ^{i+j}\,
    \frac{
      _{2}F_{1}\left[ \left\{ 3+i+j,i+j-2k\right\} ,\left\{
          4+i+j\right\} ,
        \frac{
          r_{\mathrm{N}}}{R}\right] }{3+i+j}\right\} \, ,
  \label{V_U}
\end{eqnarray}
%
where $B \left(x; h,k\right)$ is the incomplete beta function, $_{p}F_{q}$ is the generalized hypergeometric function, and $
\left( \right) _{s}$ is the Pochhammer symbol.

%
%

\subsubsection{B-splines expansion}
\label{B_spline expansion}

Since the B-spline basis set approach has been discussed in detail elsewhere~\cite{3379,8}, here we will restrict ourselves 
to a very brief compilation of its basic expressions. Following de Boor~\cite{11}, we divide the interval of interest $[0, R]$ 
into segments whose endpoints define a knot sequence $\{t_{i}\}=1,2,\ldots ,n+k.$ 
The B--splines of the order $k$, $B_{i,k}(r)$, are defined on this knot sequence by the recurrence relation 
\begin{equation}
  B_{i,k}(r) = \frac{r-t_{i}}{t_{i+k-1}-t_{i}}B_{i,k-1}(r)
  + \frac{t_{i+k}-r}{t_{i+k}-t_{i+1}}B_{i+1,k-1}(r) \, ,
\label{bs2}
\end{equation}
where the B--splines of the first order read as
\begin{equation}
  B_{i,1}(r)=\left\{
  \begin{array}{l}
  1,\qquad t_{i}\leq r\leq t_{i+1} \\
  0,\qquad \mbox{otherwise}
\end{array}
\right. \, .
\label{bs1}
\end{equation}
Note, that in these expressions the number of knots $t_{i}$ is by k larger than the number of splines. The first and the last $k
$ knots must be equal and are defined as: $t_{1} = t_{2} = \ldots = t_{k} = 0$ and $t_{n+1} = t_{n+2} = \ldots = t_{n+k} = R$. 

%
%
\subsection{Two--photon decay rates}
\label{Dif_total_decay_rates}

In the previous Sections we have obtained the finite (discrete) basis solutions of the Dirac eigenproblem (\ref
{eigenproblem}) constructed from the B--polynomial as well as B--spline sets. Now we are ready to apply these solutions for 
studying the two--photon transitions in hydrogen--like ions. Not much has to be said about the theoretical background for 
describing such a second--order process. In the past, relativistic calculations of both, the total and the differential two--photon decay rates have been discussed in detail elsewhere ~\cite{388, 3007, 3140, 3}. Below, therefore, we will repeat just 
basic expressions, relevant for discussing the role of the finite basis--sets in the two--photon calculations.

Usually, the properties of the two--photon atomic transitions are evaluated within the framework of the second--order 
perturbation theory. This theory provides the following expression for the differential in 
energy decay rate ~\cite{12},
\begin{eqnarray}
  \fl
  \frac{dw}{d\omega _{1}} &=&\frac{\omega _{1}\omega _{2}}{(2\pi )^{3}c^{2}}
  \left\vert \sum_{\nu} \left( \frac{\left\langle f\left\vert A_{2}^{\ast }\right\vert
          \nu \right\rangle \left\langle \nu \left\vert A_{1}^{\ast }\right\vert
          i\right\rangle }{E_{\nu}-E_{i}+\omega _{1}}\right.\right. 
  \left. \left.+\frac{\left\langle f\left\vert A_{1}^{\ast }\right\vert
          \nu \right\rangle \left\langle \nu \left\vert A_{2}^{\ast }\right\vert
          i\right\rangle }{E_{\nu}-E_{i}+\omega _{2}}\right)
  \right\vert ^{2}d\Omega _{1}d\Omega_{2} \, ,
  \label{eq_gera}
\end{eqnarray}
where $\omega _{j}$ is the frequency and $d\Omega _{j}$ is the element of solid angle of the $j^{th}$ photon. In this 
expression, moreover, $\ketm{i} \equiv \ketm{n_i j_i \mu_i}$, $\ketm{\nu} \equiv \ketm{(\epsilon_\nu) n_\nu j_\nu \mu_\nu}$  $
\ketm{f} \equiv \ketm{n_f j_f \mu_f}$ denote solutions of the Dirac's equation for the initial, intermediate and final ionic states 
respectively, while $E_i$, $E_\nu$ and $E_f$ are the corresponding one--particle energies. Because of energy 
conservation, $E_i$ and $E_f$ are related to the energies $\omega_{1,2}$ of the emitted photons by 
\begin{equation}
  E_{f}-E_{i}=  \omega _{1}+ \omega _{2} = \omega_{\mathrm t} \, .
  \label{con_ener}
\end{equation}

For a photon plane-wave with propagation vector $\mbox{\boldmath$k$}_j$ and polarization vector $\hat{\mbox{\boldmath
$e$}}_j$ ($\hat{\mbox{\boldmath$e$}}_j \cdot \mbox{\boldmath$k$}_j$=0), the operators $A_{j}^{*}$ in Eq.~(\ref{eq_gera}) 
are given by
\begin{equation}
       A_{j}^{*}= \mbox{\boldmath$\alpha$} \cdot
       \left( \hat{\mbox{\boldmath$e$}}_j +G \hat{\mbox{\boldmath$k$}}_j \right)
       e^{-i \mbox{\boldmath$k$}_j \cdot \mbox{\boldmath$r$}}
       - G e^{-i \mbox{\boldmath$k$}_j \cdot \mbox{\boldmath$r$}}
       \label{2.3}
\end{equation}
where $\mbox{\boldmath$\alpha$}$ is the vector of Dirac matrices and $G$ is an arbitrary gauge parameter.  Among the 
large variety of possible gauges, Grant \cite{6} showed that there are two values of $G$ which are of particular utility 
because they lead to well--known nonrelativistic operators. If $G=0$, one has the so--called Coulomb gauge, or velocity 
gauge, which leads to the dipole velocity form in the nonrelativistic limit.  If $G=\left[ \left( L+1\right) /L\right]^{1/2}$, for 
example, $G=\sqrt{2}$ for $E1$ transitions ($L=1$), one obtains a nonrelativistic expression which reduces to the dipole 
length form of the transition operator. From the general requirement of gauge invariance the final results must be 
independent of $G$. The
gauge invariance of the two--photon relativistic calculations was studied by Goldman and Drake ~\cite{3} and by Santos 
\etal ~\cite{388}.

So far, we have discussed the general expressions for the two--photon transition rates that are differential in the energy $
\omega_1$ of one of the photons. By performing an integration over this energy one may easily obtain the \textit{total} rate 
\begin{equation}
       w_{\rm tot} = \int\limits_{0}^{\omega_t} \frac{dw}{d\omega _{1}} \, {\rm d} \omega_1 ,
       \label{total_rate}
\end{equation}
which is directly related to the lifetime of a particular excited state against the two--photon decay. In this expression, we 
introduce the transition energy as $\omega_t = \omega_1 + \omega_2$.

The evaluation of the matrix elements $\left\langle f\left\vert A_{j}^{\ast }\right\vert \nu \right\rangle$ and $\left\langle 
\nu\left\vert A_{j}^{\ast }\right\vert i \right\rangle$ and, hence, of the differential (\ref{eq_gera}) as well as total (\ref{total_rate}) 
rates involves the radial integrals \cite{17}
\begin{equation}
  I_{L}^{\pm }(\omega )=\int_{0}^{\infty }\left[ P_{f}(r) Q_{i}(r) \pm
    Q_{f}(r) P_{i}(r) \right] \,j_{L}\left( \frac{\omega r}{c}\right) dr,
  \label{rri6}
\end{equation}
and
\begin{equation}
  J_{L}(\omega )=\int_{0}^{\infty }\left[ P_{f}(r) P_{i}(r) +Q_{f}(r) Q_{i}(r) \right]
  \,j_{L}\left( \frac{\omega r}{c}\right) dr,
  \label{rri8}
\end{equation}
where $j_{L}\left( x \right)$ is the spherical Bessel function of the first kind ~\cite{3297}. By making use of a a finite piecewise 
polynomials basis--set to describe the large and small radial components of the Dirac wavefunctions [cf. Eq.~(\ref{rde4})], 
both, $I_{L}^{\pm }(\omega )$ and $J_{L}(\omega )$ can be reduced to a linear combination of the integrals 
\begin{equation}
  \int_{0}^{\infty } F^{L,S}(r) F^{L,S}(r) j_{L}\left( \frac{\omega r}{c}\right)
  dr = \sum_{i=1}^{n}\sum_{j=1}^{n} f^{L,S}_{i} f^{L,S}_{j}(j_{L})_{ij} \, .
  \label{integral_jl}
\end{equation}
Here, for the sake of shortness, we denote $F^{L}(r) = P(r)$ and $F^{S}(r) = Q(r)$ as well as $f^{L}_i = p_i$ and $f^{S}_i = 
q_i$, and the matrix elements $(j_{L})_{ij}$ are defined by
\begin{equation}
    (j_{L})_{ij}=\int_{0}^{\infty }B_{i}(r)B_{j}(r)j_{L}\left(
    \frac{\omega r}{c} \right) dr \, .
    \label{j_matrix_element}
\end{equation}
These radial matrix elements are the ``building blocks'' used to evaluate two--photon decay rates. An efficient and fast evaluation of these matrix elements is crucial for studying transitions not only in hydrogen--like but also in many--electron ions. Such an evaluation can be easily performed if we employ the B--polynomials as basis--set functions. Inserting Eq.~(\ref{B_poly}) into (\ref{j_matrix_element}) we find that $(j_{L})_{ij}$ matrix elements are given \textit{analytically} by
%
\begin{eqnarray}
  (j_{L})_{ij} &=& R\left( \frac{\omega R}{c}\right)  ^{L}
  \left(\begin{array}{c} k \\ i \end{array} \right)
  \left(\begin{array}{c} k \\ j \end{array} \right)
  \frac{\pi \left( i+j+L\right) !\left( 2k-i-j\right) !}{2^{2(L+k+1)}}
  \nonumber \\
  && \times \ \  _{2}\widetilde{F}_{3}\left[ \left\{
      \frac{i+j+L+1}{2},\frac{i+j+L+2}{2}\right\} \right. ,
  \nonumber \\
&&  \left. \left\{ \frac{2L+3}{2},\frac{2k+L+2}{2},\frac{2k+L+3}{2}\right\} ,-\left(
      \frac{\omega R}{2c}\right) ^{2}\right] ,
  \label{integral_jLanal}
\end{eqnarray}
%
where $_{p}\widetilde{F}_{q}$ is the regularized generalized hypergeometric function,
\begin{eqnarray}
_{p}\widetilde{F}_{q}\left[ \left\{ a_{1},...,a_{p}\right\} ,\left\{
b_{1},...,b_{q}\right\} ,x\right] &=&\frac{1}{\Gamma \left( b_{1}\right)
...\Gamma \left( b_{q}\right) } 
\, \sum_{s=0}^{\infty }\frac{\left( a_{1}\right)
_{s}..\left( a_{p}\right) _{s}}{\left( b_{1}\right) _{s}..\left(
b_{q}\right) _{s}}\frac{x^{s}}{s!},
\end{eqnarray}
and $\Gamma(x)$ is the gamma function.

%
%
\section{Results and discussion}
\label{Results_discussion}
\subsection{Determination of the optimal set of parameters}

Having discussed the application of the finite basis solutions constructed from the B--spline and B--polynomials sets for the two--photon relativistic studies, we will employ now these solutions to analyze the properties of the $2s_{1/2} \rightarrow 1s_{1/2}$ transition in the hydrogen--like ions. Before we start an analysis, however, we must find the \textit{optimal set of parameters}, such as the number of the basis functions or the size $R$ of the cavity radius, to be used in these calculations. 

To determine the optimal set for the B-polynomial approach, we consider first the transition energy $\omega_{\rm t}$ [Eq. (\ref{con_ener})]. Accurate computation of this energy is crucial since both the differential [Eq. (\ref{eq_gera})] and total [Eq. (\ref{total_rate})] decay rates depend quadratically on it.
%
%
%
Detailed calculations of this energy value were carried out by making use of B--polynomial basis sets for various numbers of polynomials, $n_{\mathrm{BP}}$, and of the cavity radius, $R$. For the particular case of the point--like  nucleus, results of our B--polynomial calculations have been compared against the predictions of the well--known expression for the Dirac bound-state energies (see, e.g.,  Ref. \cite{215}) 
%
\begin{equation}
  \varepsilon^{\mathrm{Exact}}_{n} = c^2 \left[
    1 + \frac{\left( Z \alpha \right)^2}
    {\left[ n - \left| \kappa \right| +
        \sqrt{\kappa^2 -
          \left( Z \alpha \right)^2} \right]^2} 
  \right]^{-1/2} - c^2 \, ,
  \label{energy_eq_dirac}
\end{equation}
where $n$ is the principal quantum number and $\kappa$ is the angular quantum momentum of electron [cf. Eq.~(\ref{pkappa})]. 

Apart from the energy values $\varepsilon^{\mathrm{BP}}_{n}$, evaluation of two--photon transition rates (\ref{eq_gera}) and (\ref{total_rate}) requires also detailed knowledge of the atomic wavefunctions. As usual in atomic structure calculations, an indication for the completeness and quality of the basis set can be obtained from the comparison of the results obtained within two different gauges [cf. Eq.~(\ref{2.3})]. In this way, detailed calculations of the total decay rates for the leading two-photon $2E1$  $2s_{1/2}\rightarrow 1s_{1/2}$ channel have been performed, in both length and velocity gauges. The gauge invariance condition together with the the energy $\omega_{t}$  accuracy determine the optimal set of parameters.

%
%
%

The relative difference $\Delta_{\omega_{t}} = | (\omega^{\mathrm{Exact}}_{t} - \omega^{\mathrm{BP}}_{t})/\omega^{\mathrm{Exact}}_{t} |$ between the exact $\omega^{\mathrm{Exact}}_{t}$ solution and the basis set value $\omega^{\mathrm{BP}}_{t}$ is presented in Fig.~\ref{merged_energies} as a function of $n_{\mathrm{BP}}$ for $Z$ equal to   1, 40 and 92.
%
%
We display in Fig.~\ref{merged_gauge}  the relative difference between the length   and velocity gauge decay rate values of the $2E1$ $2s_{1/2}\rightarrow 1s_{1/2}$, $\Delta_{\mathrm{l-v}}$, as function   of the number of B-Polynomials, $n_{\mathrm{BP}}$, for the same values of the atomic number $Z$.

%
 As seen from Figs. ~\ref{merged_energies}  and ~\ref{merged_gauge}, the optimal $n_{\mathrm{BP}}$ value for  $Z=1$ is greater than 34 and the optimal $R$ is 50.  For  $Z=40$ the parameters are $n_{\mathrm{BP}}>40$ and $R=1$, and for $Z=92$ are $n_{\mathrm{BP}}>40$ and $R=0.25$.

%


Working in double precision, we noticed a loss of numerical significance in the results for  $n_{\mathrm{BP}}> 23$ due to the LAPACK 3.3.0 routines used for the solution of the generalized eigenvalue problem.
%
%
We detected that this is due to the difficulty of these routines, namely the  DSYGVX routine, to deal with diagonally dominant matrices that have very large diagonal elements and very small off diagonal elements. 
%
This problem was overcome by using quadruple precision for the evaluation of the matrix elements and compiling the LAPACK subroutine in quadruple precision as well, using the compiler auto-doubling option.
This situation is illustrated in Fig. \ref{Z_1_R40_Inv}, where it is plotted the relative  difference $\Delta_{\mathrm{l-v}}$ for $Z=1$ and $R=40$, obtained in double and quadruple precision.
%
%

Nevertheless, we should emphasize that even with a low $n_{\mathrm{BP}}$ value, such as 20, and working in double precision, we get a gauge invariance in the two-photon rate less than $10^{-16}$. 

Before turning to the second--order relativistic calculations performed within the finite--basis--set approaches, let us recall that apart from the B--polynomials, we used also  the ``standard'' B--spline solutions. For these solutions, we adopted the following parameters described in Ref.~\cite{3007}: $k=9$, $n_{\mathrm{BS}}=60$, and $R=60$ a.u.  Moreover, the integration over the photon frequency in Eq.~(\ref{total_rate}) has been performed using a 15--point Gauss--Legendre algorithm. 
%


\subsection{Second--order relativistic calculations}

The ``optimal'' sets of parameters for both, the B--spline and B--polynomial approaches, determined in the previous section 
will be used below for the computation  of the two--photon transition rates in quadruple precision. However, first let us briefly return to the energy 
calculations performed in these two approaches. In Table ~\ref{Energies} we report the relative differences between the 
computed energies of $n s_{1/2}$ states of neutral hydrogen and the corresponding exact values $\varepsilon^{\mathrm
{Exact}}_{n}$ defined by Eq.~(\ref{energy_eq_dirac}). Here, $\Delta E^{\mathrm{B-Pol}}_{\mathrm{TW}}$ and $\Delta E^
{\mathrm{B-splines}}_{\mathrm{TW}}$ represent the differences obtained by employing the B--polynomial and B--spline 
sets, correspondingly. Moreover, $\Delta E^{\mathrm{B-Pol}}_{\mathrm{BP}}$ denotes the relative difference calculated with 
the B--polynomial values by Bhatti and Perger ~\cite{3363}. 
%
The B--polynomial relative differences are very similar with the B--Spline relative differences, for $n>2$. For $n=1,2$, the former method achieved an excellent 12 digits agreement with the exact result, two orders of magnitude better than the B-Splines.


In Table ~\ref{Multipole} we display the most significant multipole contributions to the $2s_{1/2}\rightarrow 1s_{1/2}$ two--photon total decay rate [Eq. (\ref{total_rate})] of neutral hydrogen. Again, calculations have been performed within two different  gauges and by employing B--polynomial as well as B--spline basis sets. Multipoles with only magnetic components such as $2M1$ only have velocity gauge. 
As seen in this Table, both approaches yield almost identical results (with the relative error of less than 2 $\times $ 10$^{-7}$) for all multipole decay channels. 
Furthermore, one may observe a remarkable agreement (smaller than $10^{-25}$) between the values obtained in the length and velocity gauges for both basis sets. 

It should be emphasized that with the B-Polynomials approach, 
that employs \textit{analytical} evaluation of the second--order matrix elements [cf. Eqs.~(\ref{rri6})-- (\ref{integral_jLanal})] and uses a smaller  optimal $n_{\mathrm{BP}}$, the computation time required for the determination of each multipole contribution 
is, in quadruple precision, about two and half times smaller than the time required by the B--Splines methods. 
%
}
%
These results clearly indicate that the finite--basis--set approach based on B--polynomial solutions provides an alternative tool for studying the two--photon  transitions. 

Besides the neutral hydrogen, extensive test of the B--polynomial approach has been performed also for the two--photon 
transitions in medium-- and high--$Z$ hydrogen--like ions. In Table~\ref{tab_continuum} we display, for example, the total 
rates of the leading, $2E1$ $2s_{1/2}\rightarrow 1s_{1/2}$ channel for selected values of the nuclear charge $Z$. Here we 
used both, B--spline and B--polynomial approaches to carry out the intermediate--state summation in Eq.~(\ref{eq_gera}) 
over the complete Dirac's spectrum ($W^{T}$) as well as over the positive-- ($W^{+}$) and negative--energy ($W^{-}$) 
solutions only. The role of Dirac's continua in relativistic second--order calculations has been the subject of recent 
theoretical investigations ~\cite{3140,1418}
 partially because of their impact on (future) many--body studies. As seen from  Table~\ref{tab_continuum}
 , the good agreement between the predictions obtained within the B--polynomial and B--spline approaches can be  found for the ions along the entire isoelectronic sequence. Moreover, both our calculations show a perfect agreement between gauges, which is better than the one reported by Labzowsky \etal~\cite{1418}. 
 
 Again, the presented calculations validate the B--Polynomial basis set as a useful and appropriate basis set to study two--photon decays in atomic systems, and various atomic calculations that involve a summation over the Dirac spectrum.
%
%
\section{Conclusions}
\label{Conclusions}

In the present work we have investigated, for the first time, the efficiency and accuracy of the B--Polynomial basis set  for studying the two--photon transitions in hydrogen--like systems. Based on the finite--basis--set approach, the 
generalized eigenproblem in Eq. (\ref{rde5}) was solved in order to provide the eigenvalues and eigenfunctions, 
which were successfully used for the calculations of the two--photon decay rates. We took advantage of the B--Polynomial 
properties and derived fully relativistic \textit{analytical} expressions for the two--photon rates both, within the pointlike-- and 
finite--nucleus models, instead of employing numerical methods.

In order to illustrate the application of the B--polynomial method and to verify its accuracy, we have performed detailed calculations of the total rates  for the $2s_{1/2}\rightarrow 1s_{1/2}$ two--photon transition in neutral hydrogen and  hydrogen--like ions. Results of these calculations have been compared with the predictions of the well--established B--spline approach. While the perfect agreement between the results of both, B--spline and B--polynomial approximations was observed along the entire isoelectronic sequence, the B--polynomial calculations were found to be much less computationally demanding. 

It was noticed that if we consider a basis set with more than 23 B-Polynomials, we need to work in quadruple precision due to the LAPACK 3.3.0 routines limitation in dealing with matrices that have very large diagonal matrix elements and very small off-diagonal values. 

We conclude that the B--Polynomial basis sets are suitable for investigation of the two--photon transitions with the great advantage of enabling analytical expressions for the involved integrals, which speed up the calculations. 

The presented procedure may be easily extended to calculate two--photon transitions in \textit{many--electron} ions or atoms, as well as to the quantum electrodynamics (QED) calculations that usually require the summation over the complete Dirac spectrum.

%
%
\ack
%
%
 This research was supported in part by FCT, by the French-Portuguese  collaboration (PESSOA Program, Contract n$^\mathrm{o}$ 441.00),   by the Ac{\c c}{\~o}es Integradas Luso-Francesas (Contract n$^\mathrm{o}$ F-11/09) and  by the Programme Hubert Curien and Ac{\c c}{\~o}es Integradas Luso-Alem{\~a}s (Contract n$^\mathrm{o}$ 20022VB and A-19/09).  The work of A.S. was supported by the  Helmholtz Gemeinschaft (Nachwuchsgruppe VH--NG--421) and by  the Deutscher Akademischer Austauschdienst (DAAD) under  the Project No. 0813006. Laboratoire  Kastler Brossel is ``Unit\'e Mixte de Recherche du CNRS, de l' ENS et de l'UPMC n$^{\circ}$ 8552''. P. Indelicato acknowledges the support of the Helmholtz Allianz Program of the Helmholtz  Association, contract HA-216 "Extremes of Density and Temperature:   Cosmic Matter in the Laboratory". P. Amaro acknowledges the support  of the FCT, contract SFRH/BD/37404/2007.

%

%


\section*{References}

\bibliographystyle{unsrt}
\bibliography{jps}

%
\newpage


%
%
%
%

\newpage
%
%

\begin{table}										
  \caption{\label{Energies} Relative differences between the computed energy eigenvalues
    of hydrogen ($s$ states) using B-polynomials and B-splines and the
    exact results of the Coulomb-field Dirac equation,
    $E^{\mathrm{Exact}}$, given by Eq. (\ref{energy_eq_dirac}).
    $\Delta E^{\mathrm{B-Pol}}_{\mathrm{TW}}$
    and $\Delta
    E^{\mathrm{B-splines}}_{\mathrm{TW}}$ represent, respectively, the
    relative differences calculated in this work with
    the B-polynomials
    and
    with the B-splines. $\Delta E^{\mathrm{B-Pol}}_{\mathrm{BP}}$ denote
    the relative difference obtained with the B-polynomials values by
    Bhatti and Perger ~\cite{3363}. Powers of ten are given in
    parentheses. } 						
\begin{indented}
  \lineup
\item[]\begin{tabular}{@{}llllllll}
\br
$n$	& &	$E^{\mathrm{Exact}}$ 	& &		$\Delta E^{\mathrm{B-Pol}}_{\mathrm{TW}}$	&	$\Delta E^{\mathrm{B-
splines}}_{\mathrm{TW}}$	&	$\Delta E^{\mathrm{B-Pol}}_{\mathrm{BP}}$	\\
\mr
\\
1	& &	-0.5000066565953603	& &		$1.6 (-12)$	&	$3.8 (-10)$	&	$2.7 (-12)$	\\
2	& &	-0.12500208018900594	& &		$1.4 (-12)$	&	$1.4 (-10)$	&	$1.5 (-8)$	\\
3	& &	-0.05555629517766647	& &		$1.7 (-7)$		&	$3.4 (-9)$	&		$1.8 (-3)$	\\
4	& &	-0.03125033803007682	& &		$1.4 (-3)$		&	$5.9 (-5)$	&		$2.3 (-1)$	\\
\br
\end{tabular}											
\end{indented}
\end{table}											

%
%
\pagebreak


\begin{table}
\caption{\label{Multipole} Multipole contributions (in s$^{-1}$) of the
  $2s_{1/2}\rightarrow 1s_{1/2}$ two-photon decay for $Z=1$.
The relativistic calculations have been performed within the
    velocity and length gauges, using the B-polynomials and B-splines
    basis sets.
$\Delta_{\mathrm{l-v}}$ stands for the
    relative difference between the length  and velocity gauge
    values.
 Powers of ten are given in parentheses. }						
\begin{indented}
  \lineup
\item[]\begin{tabular}{@{}llllll}
\br
Multipoles & \multicolumn{5}{c}{Contribution (s$^{-1}$)} \\
\cline{2-6} \\
& \multicolumn{2}{c}{B-Polynomials} && \multicolumn{2}{c}{B-Splines} \\
\cline{2-3}  \cline{5-6}  \\
& \multicolumn{1}{c}{lenght gauge} & \multicolumn{1}{c}{$\Delta_{\mathrm{l-v}}$} &&
\multicolumn{1}{c}{length gauge} &
\multicolumn{1}{c}{$\Delta_{\mathrm{l-v}}$} \\
\mr
\\
$2E1$  & $8.2290591586  $      & $<1.0(-26)$            && $8.2290591509 $   & $<1.0(-15)$ \\
$E1M2$ & $2.5371807735 (-10)$ & $<1.0(-25)$            && $2.5371807635 (-10)$ & $<1.0(-15)$ \\
$2M1$  & $1.3803580496  (-11)$ & \multicolumn{1}{c}{--} && $1.3803580473 (-11)$ & \multicolumn{1}{c}{--} \\
$2E2$  & $4.9072289232  (-12)$ & $<1.0(-34)$            && $4.9072289165 (-12)$ & $<1.0(-14)$ \\
$2M2$  & $3.0693510074  (-22)$ & \multicolumn{1}{c}{--} && $3.0693509833 (-22)$ & \multicolumn{1}{c}{--} \\
$E2M1$  & $1.6393565197  (-23)$ & $<1.0(-34)$   && $1.6397413530 (-23)$ & $<1.0(-4)$ \\
Total        & $8.2290591589$ &				&&	8.2290591512			  &    \\
\br
\end{tabular}
\end{indented}
\end{table}

%
%


\begin{table}
  \caption{\label{tab_continuum}
    Total two-photon $2E1$   $2s_{1/2}\rightarrow 1s_{1/2}$
    decay rates (in s$^{-1}$) for selected values of the nuclear charge
    $Z$. The relativistic calculations have been performed within the
    velocity and length gauges, using the B-polynomials and B-splines
    basis sets, and by carrying out intermediate-state
    summation over  the complete Dirac's spectrum ($W^{\mathrm{T}}$) as well as
    over the positive-  ($W^{+}$) and negative-energy
    ($W^{-}$) solutions only. $\Delta_{\mathrm{l-v}}$ stands for the
    relative difference between the length  and velocity gauge
    values, and LSS denotes the values calculated by Labzowsky
    \etal~\cite{1418}. Powers of ten are given in parentheses. 
}
\begin{indented}
  \lineup
\item[]\begin{tabular}{@{}llllllllll}
\br
\\
$Z$ & \multicolumn{9}{c}{Decay rates (s$^{-1}$)} \\
\cline{2-10} \\
&& \multicolumn{2}{c}{B-Polynomials} && \multicolumn{2}{c}{B-Splines}  && \multicolumn{2}{c}{LSS}\\
\cline{3-4}  \cline{6-7} \cline{9-10}  \\
&& \multicolumn{1}{c}{lenght gauge} & \multicolumn{1}{c}{$\Delta_{\mathrm{l-v}}$} &&
\multicolumn{1}{c}{length gauge} & \multicolumn{1}{c}{$\Delta_{\mathrm{l-v}}$}&&
\multicolumn{1}{c}{length gauge} &
\multicolumn{1}{c}{$\Delta_{\mathrm{l-v}}$} \\
\mr
\\
1  & $W^{+}$ & $8.22861$   &       && $8.22861$      &      &&	$	8.2206	$	&	$		$	\\
   & $W^{-}$ & $2.49477(-8)$   &       && $2.49481(-8)$      &      &&	$	3.9975(-22)	$	&	$		$	\\
   & $W^{\mathrm{T}}$ & $8.22906 $     & $<1.0(-26)$  && $8.22906 $ & $<1.0(-15)$ &&	$	8.2207	$	&	$	<1.0
(-4)	$	\\



40 & $W^{+}$ & $2.96130(+10)$   &       && $2.96124(+10)$      &       &&	$	3.1953(+10)	$	&	$		$	\\
   & $W^{-}$ & $2.10270(+8)$   &       && $2.10250(+8)$      &      &&	$	5.8284	$	&	$		$	\\
   & $W^{\mathrm{T}}$ & $3.19862(+10)$ & $<1.0(-13)$  && $3.19858(+10)$ & $<1.0(-15)$ &&	$	3.1953(+10)	$
	&	$	<1.0(-4)	$	\\



92 & $W^{+}$ & $2.90482(+12)$   &       && $2.90409(+12)$      &      &&	$	3.8230(+12)	$	&	$		$	\\
   & $W^{-}$ & $6.85553(+11)$   &       && $6.80648(+11)$      &      &&	$	1.2851(+5)	$	&	$		$	\\
   & $W^{\mathrm{T}}$ & $3.825839(+12)$ & $<1.0(-9)$  && $3.82555(+12)$ & $<1.0(-15)$ &&	$	3.8216(+12)	$	&
	$	<1.0(-5)	$	\\
\br
\end{tabular}
\end{indented}
\end{table}

%
%
\Figures

\includegraphics[width=17cm]{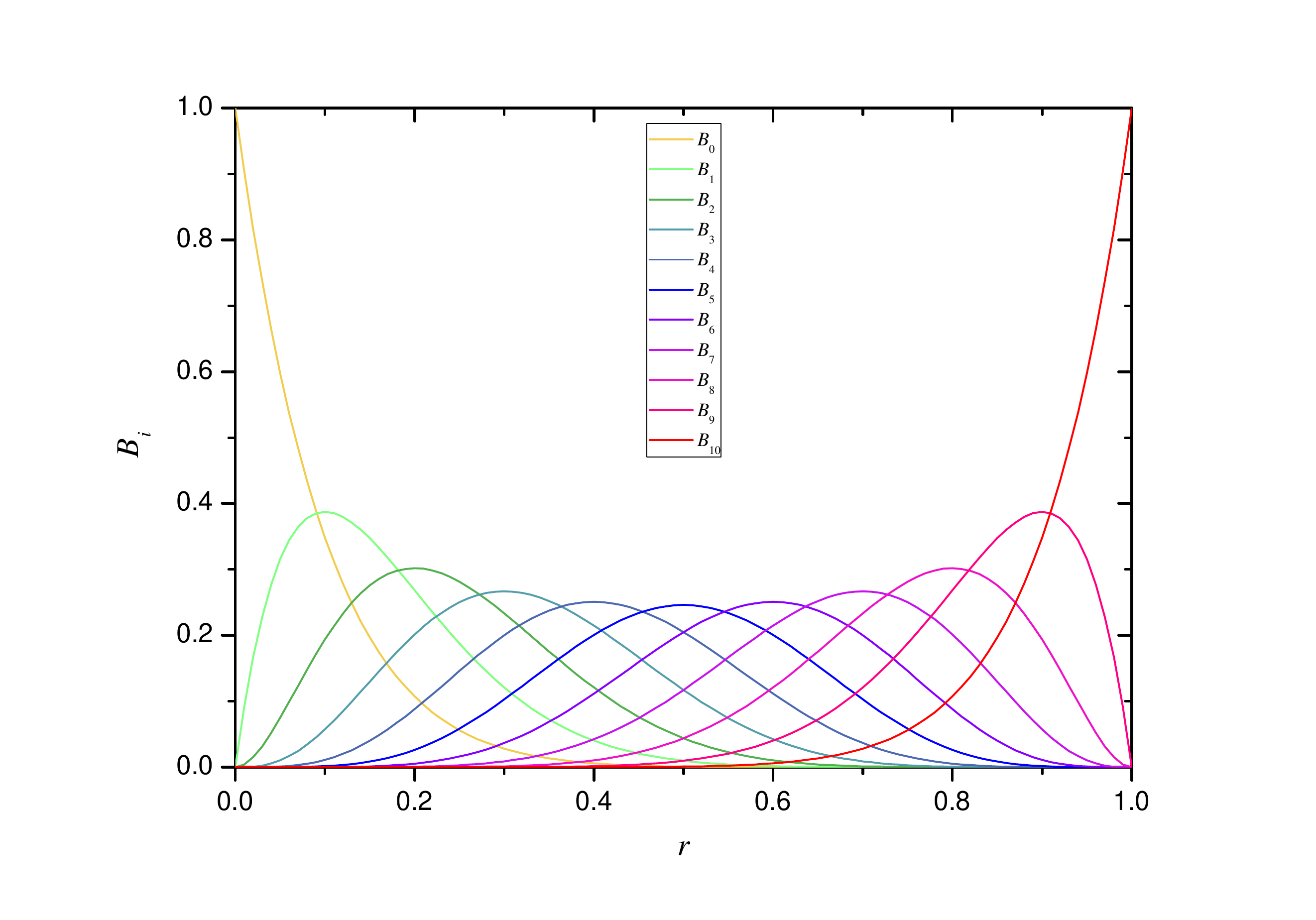}

\Figure{\label{fig_bpolys} The set of 11 B-Polynomials of degree 10 are shown in the interval $\left[0,1 \right]$. The quantities are dimensionless.}
%
%
%

\includegraphics[width=17cm]{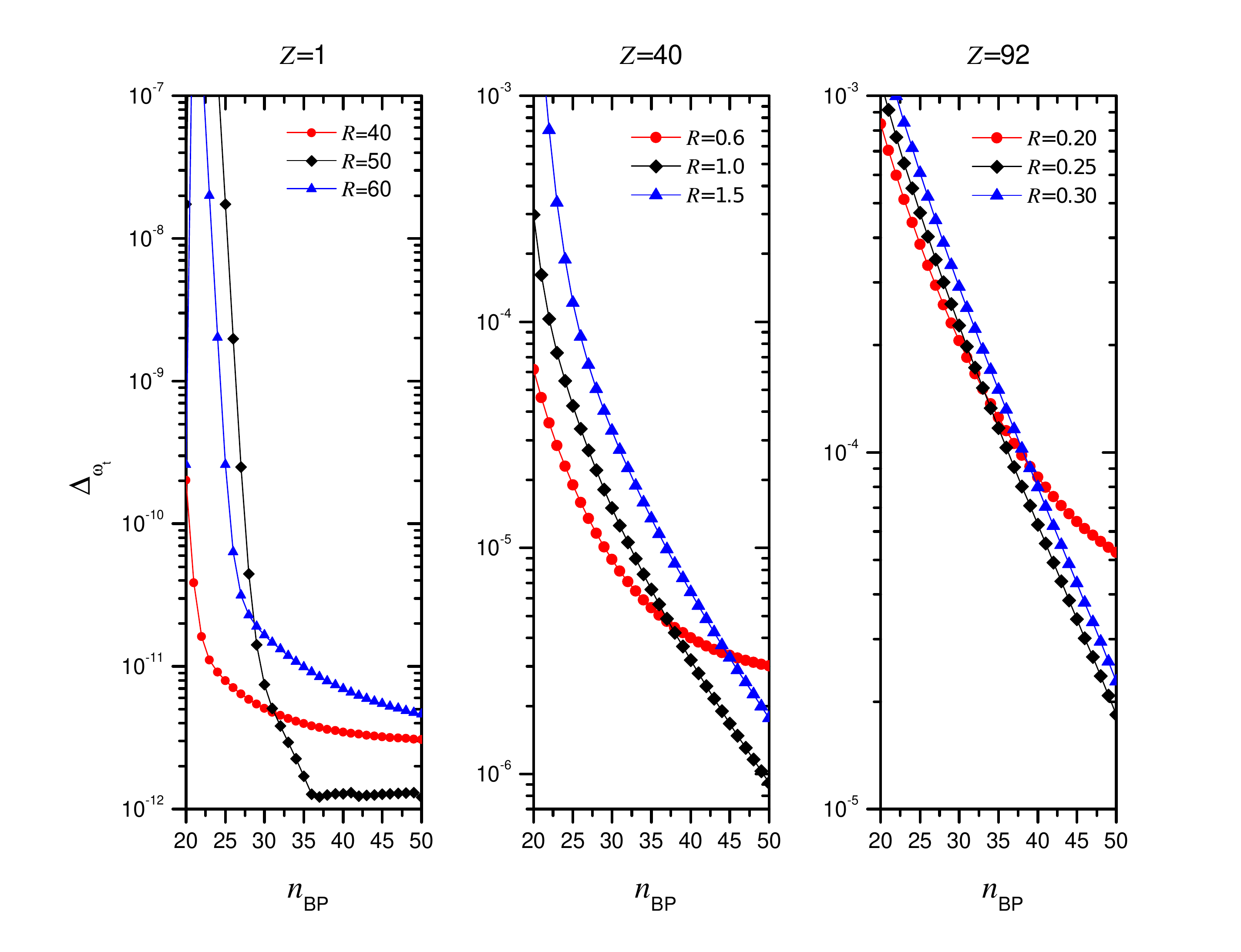}

\Figure{\label{merged_energies} Relative difference $\Delta_{\omega {\mathrm t}}$ between the   computed $\omega_{\mathrm t}$  energy, given by Eq.~\ref{con_ener}, and the value   $\omega ^{\mathrm{Exact}}_{t}=E^{\mathrm{Exact}}_{2s}-E^{\mathrm{Exact}}_{1s}$  as function of the number of B-Polynomials, $n_{\mathrm{BP}}$, for   $Z$ equal to 1, 40 and 92. Here, $E^{\mathrm{Exact}}_{n}$ are the   exact solutions of the Coulomb-field Dirac equation given by   Eq.~(\ref{energy_eq_dirac}). }
%
%

\includegraphics[width=17cm]{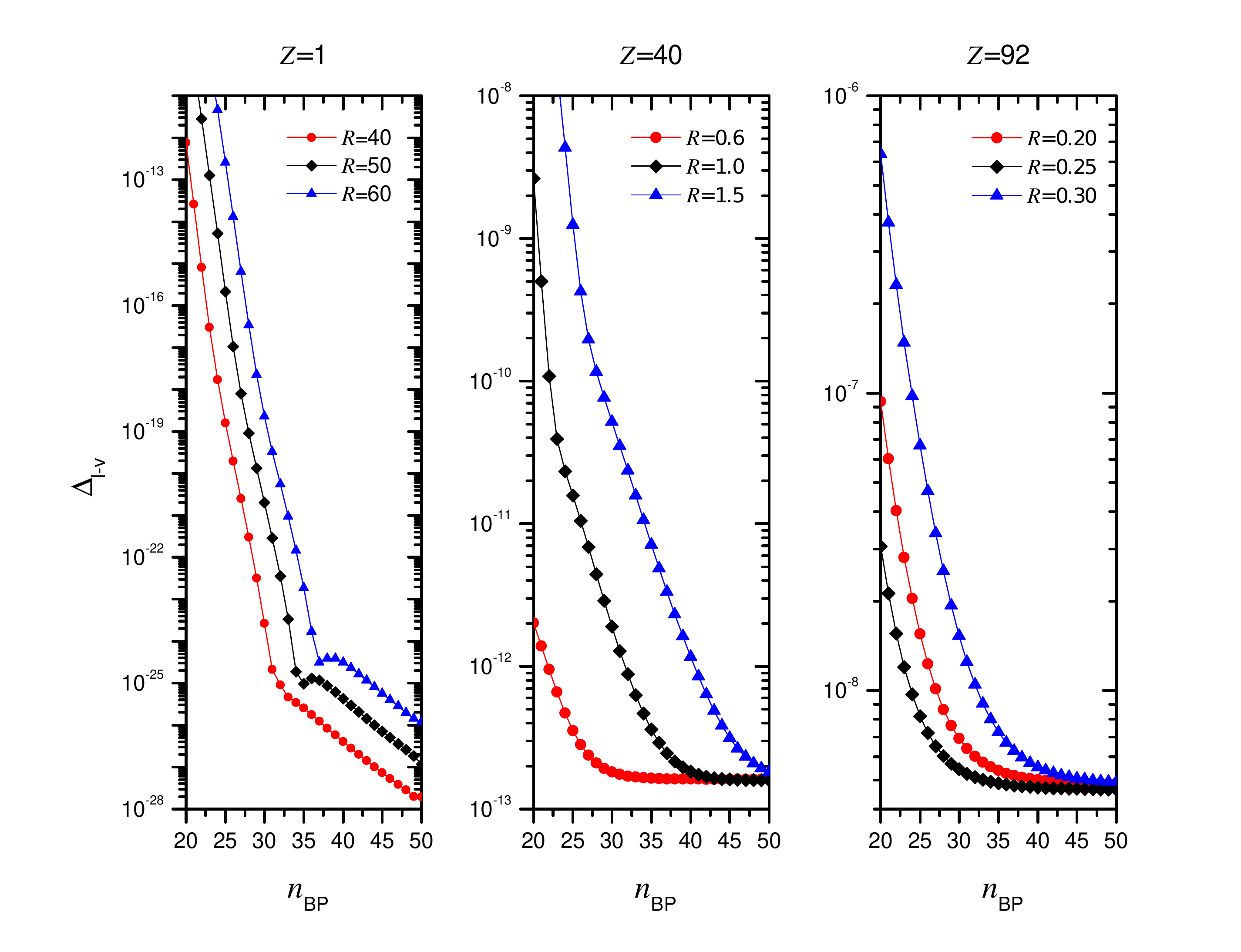}

\Figure{\label{merged_gauge} Relative difference between the length   and velocity gauge decay-rate values of the $2E1$ $2s_{1/2}\rightarrow 1s_{1/2}$, $\Delta_{\mathrm{l-v}}$, as function   of the number of B-Polynomials, $n_{\mathrm{BP}}$, for $Z$ equal to   1, 40 and 92. }
%
%

\includegraphics[width=17cm]{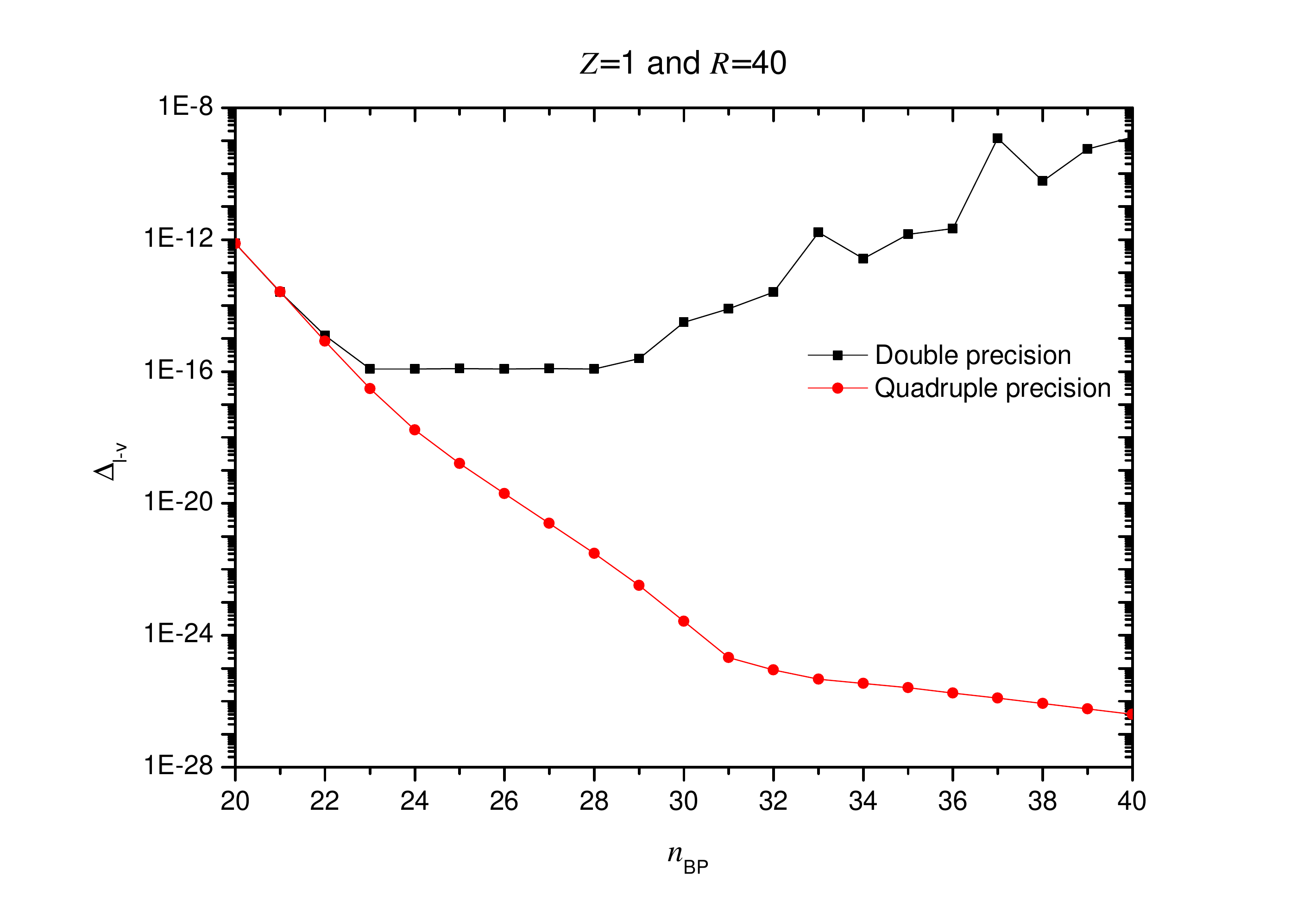}

\Figure{\label{Z_1_R40_Inv} Relative difference between the length  and velocity gauge decay-rate of the $2E1$ $2s_{1/2}\rightarrow 1s_{1/2}$, $\Delta_{\mathrm{l-v}}$, for $Z=1$ and $R=40$, obtained in double and quadruple precision.}

%
\end{document}